\documentclass[prl,nofootinbib,twocolumn,english]{revtex4}
\usepackage{amsmath}
\usepackage{amsfonts,color}
\usepackage{tensor}
\usepackage{amssymb,float}
\usepackage[utf8]{inputenc}
\usepackage{color}
\usepackage[unicode=true,pdfusetitle,
 bookmarks=true,bookmarksnumbered=true,bookmarksopen=true,bookmarksopenlevel=3,
 breaklinks=false,pdfborder={0 0 1},backref=false,colorlinks=true]
 {hyperref}
\usepackage{enumitem}
\usepackage{graphicx}
\usepackage{appendix}
\usepackage{hyperref}
\usepackage{lipsum}

\makeatletter
\long\def\dddddot#1{%
  {\mathop {#1}\limits ^{\vbox to-1.4\ex@ {\kern -\tw@ \ex@ \hbox {\normalfont .....}\vss }}}%
}
\long\def\multidots#1#2{%
  \count@=0
  {{\mathop {#2}\limits ^{\vbox to-1.4\ex@ {\kern -\tw@ \ex@ \hbox {\normalfont %
  \loop%
  \ifnum#1>\count@%
  .%
  \advance\count@ by1%
  \repeat%
  }\vss }}}}%
}
\makeatother

\begin{document}

\title{The first non-static inhomogeneous exact solution in $f(T,B)$ gravity}

\author{Sebastián N\'ajera}
\email{najera.sebastian@ciencias.unam.mx}
\affiliation{Instituto de Ciencias Nucleares, Universidad Nacional Aut\'onoma de M\'exico, 
Circuito Exterior C.U., A.P. 70-543, M\'exico D.F. 04510, M\'exico.}

\author{Aram Aguilar}
\email{armandoaram$\_$94@hotmail.com}
\affiliation{Instituto de Ciencias Nucleares, Universidad Nacional Aut\'onoma de M\'exico, 
Circuito Exterior C.U., A.P. 70-543, M\'exico D.F. 04510, M\'exico.}

\author{Celia Escamilla-Rivera}
\email{celia.escamilla@nucleares.unam.mx}
\affiliation{Instituto de Ciencias Nucleares, Universidad Nacional Aut\'onoma de M\'exico, 
Circuito Exterior C.U., A.P. 70-543, M\'exico D.F. 04510, M\'exico.}

\author{Roberto A. Sussman}
\email{sussman@nucleares.unam.mx}
\affiliation{Instituto de Ciencias Nucleares, Universidad Nacional Aut\'onoma de M\'exico, 
Circuito Exterior C.U., A.P. 70-543, M\'exico D.F. 04510, M\'exico.}

\begin{abstract}
We examine in this paper the possibility of finding exact solutions for Teleparallel Gravity (TG) of the type of spherically symmetric Lema\^\i tre-Tolman-Bondi (LTB) dust models. We apply to the LTB metric, as obtained from the Schwarzschild solution in General Relativity, the formalism of Teleparallel Gravity in its extension to $f(T,B)$ models. An exact LTB solution is obtained that is compatible with a specific  $f(T,B)$ model that seems to be appropriate to fit observations when applied to standard spatially flat Robertson-Walker geometry.
\end{abstract}

\maketitle

%%%%%%%%%%%%%%%%%%%%%%%%%%%%%%%%%%%%%%%%%%%%
%%%%%%%%%%%%%%%%%%%%%%%%%%%%%%%%%%%%%%%%%%%%
\textit{Introduction}.- Cosmology has benefited in the last decades from an enormous increase of new observational data whose proper interpretation is still a challenging endeavor, not only because the fundamental nature of the dark sector is still an open research topic, but also from existing \textit{tensions} around the values of $H_0$ and $S_8$ and anomalies in early-time observations, all of which deserve further investigation \cite{DiValentino:2020vhf, di2021realm}.
 
There is currently a wide variety of dark energy proposals to explain the large scale accelerated cosmic expansion under the framework of General Relativity (GR) under gauge invariant cosmological perturbations on a Friedmann-Lema\^\i tre-Robertson-Walker (FLRW) background. An accelerated cosmic expansion can be obtained by assuming a positive $\Lambda$, extra \textit{dark fluids} with negative pressure, scalar fields or modifications of General Relativity (GR), among others \cite{Clifton:2011jh, ishak2019modified}.
\noindent More recently, extensions of Teleparallel Gravity (TG) \cite{Cai:2015emx, Hohmann:2019nat} have also been considered to explain cosmic acceleration, as well as to alleviate the $H_0$ tension, all this from the geometry of the theory itself without involving an extra fluid or $\Lambda$ \cite{Briffa:2020qli} and in compatibility with current observations \cite{Nunes:2018xbm,Escamilla-Rivera:2019ulu,Cai:2019bdh}.

TG is a theory of gravity in which the curvature-based description of gravity, used in GR, is enhanced by torsion, which involves replacing the Levi-Civita connection with its Weitzenböck connection analog. The name of \textit{teleparallelism} or \textit{distant parallelism} comes from the geometrical description of vectors in the Weitzenböck space-time \cite{Pereira:2013qza}, where torsion can change the direction of such vectors \cite{koivisto}. TG is a gauge theory for the translations group \cite{article}, so it is possible to represent the gravitational field with a translational gauge potential that appears within the non-trivial part of the tetrad field (and in fact this keeps torsion different from zero).

In this paper we study the compatibility between TG theories and the standard spherically symmetric Lema\^\i itre-Tolman-Bondi (LTB) models, an important class of exact inhomogeneous solutions of GR with a dust source  \cite{Krasinski,plebanski}, though the LTB metric also admits non-trivial solutions with non-zero pressure \cite{Sussman:PRD}. These solutions are very useful as \textit{toy models} to study structure formation \cite{lake2000gravitational, sussman2009quasilocal, deshingkar2001gravitational, bolejko2010structures}, as they can also be conceived as exact non-linear perturbations of a FLRW background, reducing in the linear regime to standard cosmological perturbations in the isochronous comoving gauge \cite{sussman2015spherical, Sussman:Invar,sussman2008quasi, sussman2010new}. 
 
While there is an extensive literature on LTB models in GR, in TG theories only a few articles have considered them to examine the possibility to describe dark energy dynamics without introducing a cosmological constant \cite{Cai:2015emx} and in the discussion on the problem of energy-momentum localization in TG \cite{salti2005energy}. In this context, our interest is to examine how the known properties of LTB solutions can be affected by torsion acting as a force in a possible LTB solution of TG theories. 

%%%%%%%%%%%%%%%%%%%%%%%%%%%%%%%%%%%%%%%%%%%%%%%%%%%%%%%%%%%%%%%%%
%%%%%%%%%%%%%%%%%%%%%%%%%%%%%%%%%%%%%%%%%%%%%%%%%%%%%%%%%%%%%%%%%
\bigskip

\textit{From Schwarzschild to LTB solutions: the GR approach}.- Consider the Schwarzschild spacetime in standard static coordinates\footnote{We will consider geometric units ($c = G = 1$) along the rest of the paper.} 
\begin{equation}
ds^2=-\left(1-\frac{2M_0}{r}\right)dt^2 + \left(1-\frac{2M_0}{Y}\right)^{-1} dY^2 + Y^2d\Omega^2,
\end{equation}
where $M_0$ is the Schwarzchild mass. Its corresponding radial timelike geodesics follow from: 
\begin{equation} \dot{Y}^2 = \frac{2M_0}{Y}-k,\quad \dot t = \frac{\sqrt{1-\kappa}}{1-2M_0/Y}, \quad k \equiv \frac{k_0^2}{4}-1,\label{rtdots}\end{equation}
where $k_0$ is the conserved energy in each geodesic. From Eq.~(\ref{rtdots}) we can construct a coordinate system $(\tau,r)$ for the Schwarzschild spacetime based on the world lines of observers following radial geodesics \cite{plebanski} with their proper time $\tau$ as time coordinate and $r$ acting as a continuous parameter that marks each geodesic. Since the binding energy $k$ continuously varies along the geodesic congruence in the $(\tau,r)$ plane, we have necessarily:  $k=k(r)$ and $Y=Y(\tau,r),\,\,t=t(\tau,r)$. Under the coordinate transformation $(t,Y)\to (\tau,r)$ defined by Eq.~(\ref{rtdots}) the Schwarzschild static metric takes the following time dependent form:
\begin{equation} ds^2 = -d\tau ^2+\frac{Y'^2}{1-k}\,dr^2+Y^2\,d\Omega^2,\label{ScwLTB}\end{equation}
where $k=k(r)$ and $\mathcal{A}'\equiv \left[\partial \mathcal{A}/\partial r\right]_\tau$ for all functions $\mathcal{A}$, with $Y(\tau,r)$ satisfying
\begin{eqnarray}
&&\dot{Y}^2 =\frac{2M_0}{Y}+k,\label{rdot}\\ &&
\quad \tau-\tau_B(r)=\int_{\bar{Y}=0}^Y{\frac{\sqrt{\bar{Y}}d\bar{Y}}{\sqrt{2M_0+k(r)\,\bar{Y}}}},\nonumber
\end{eqnarray}
where $\dot{\mathcal{A}}\equiv u^a \mathcal{A}_{,a}= \left[\partial \mathcal{A}/\partial \tau\right]_r$ for every function $\mathcal{A}$ and $\tau_B(r)$ an integration constant marking the value of $\tau$ for which $Y=0$ ({\it i.e.} the geodesics fall or arise from the singularity at different values of $\tau$ for each $r$). These coordinates make it easy to appreciate that the Schwarzschild radius $Y_s=2M_0$ is a regular locus.

The Schwarzschild mass can be represented in the spherical coordinates of Eq.~(\ref{ScwLTB}) as an integrated distribution $M_0 =\int_V{\rho\,d^3 x}$ with $\rho=M_0\delta^3(0)/(r^2\sin\theta)$. Substituting the distributional density for a continuous density (equivalent to consider $M_0 \, \rightarrow \, M=M(r)$) and assuming the $4$-velocity $u^a =\left[dx^a/d\tau\right]_{r}= \delta^a_\tau$, for observers comoving along the geodesics, the substitution of Eqs.~(\ref{ScwLTB})--(\ref{rdot}) in Einstein equations, $G_{ab}=8\pi \Theta_{ab}$ yields
\begin{equation} \Theta^{ab} =\rho\,u^a\,u^b=\rho\,\delta^a_\tau\,\delta^b_\tau, \quad  8\pi\rho = \frac{2M'}{Y^2\,Y'},\quad u^a=\delta^a_\tau,
\label{Tab}\end{equation}
which characterize LTB models with a dust source (in fact, Eq.~(\ref{ScwLTB}) is already the the LTB metric representation of the Schwarzschild solution). It is worth noting that Eq.~(\ref{rdot}) need not hold in modified gravity theories, for example in f(R) theories (see  \cite{sussman2017lemaitre}). Nevertheless, we will consider henceforth this evolution equation for the area radius $Y$, Eq.~(\ref{rdot}) in the derivation of Schwarzschild-LTB models in TG gravity. 

%%%%%%%%%%%%%%%%%%%%%%%%%%%%%%%%%%%%%%%%%%%%%%%%%%%%%%%%%%%%%%%%%%
%%%%%%%%%%%%%%%%%%%%%%%%%%%%%%%%%%%%%%%%%%%%%%%%%%%%%%%%%%%%%%%%%%
\bigskip 

\textit{Teleparallel Gravity solutions}.-
As a gauge theory, the gravitational interaction in TG is modeled as a force and the particles trajectories are not geodesic but force equations, with the torsion tensor playing the role of force \cite{krvsvsak2019teleparallel}. Classically speaking, it is completely equivalent to describe gravitation with a pseudo-Riemannian with signature $(-,+,+,+)$ (GR) or a Weitzenböck (TG) space-time structure \cite{Pereira4}. Moreover, by a convenient choice of the TG Lagrangian, GR and TG are equal up to a boundary term, which means that they produce identical dynamical equations \cite{Bahamonde:2019shr}. The relation between the two theories is really important in the sense that one should always look for certain TG models that in a particular limit can return GR, this is appropriate if we want to describe gravity at the level where GR is useful. In this equivalence, the geodesic equation is analog to the Lorentz force equation of electrodynamics. TG is based on the tetrad formalism the tetrads or tetrad fields are considered the dynamical variables of the system. In contrast to GR in which the metric takes the role of dynamical variable \cite{celia3}. In the Weitzenböck spacetime structure, a Weitzenböck connection $\Gamma_{\enspace \mu \nu}^{\sigma}$ is defined as
\begin{equation}\label{weitconnection}
\Gamma_{\enspace \mu \nu}^{\sigma} := E_{A}^{\enspace \sigma}\partial_{\mu}e_{\enspace \nu}^{A} + E_{A}^{\enspace \sigma}\omega_{\enspace B\mu}^{A}e_{\enspace \nu}^{B},
\end{equation}
\vspace{3mm}

\noindent where $e_{\enspace \mu}^{A}$ is the tetrad field, $E_{A}^{\enspace \sigma}$ the transpose tetrad field defined by $E_{A}^{\enspace \mu}e_{\enspace \nu}^{A}=\delta^{\mu}_{\nu}$ and $\omega_{\enspace b\mu}^{A}$ is the spin connection. Space-time indices are denoted by greek letters while capital letters stand for tangent space indices. In all this work, quantities with/without a white circle upon them are calculated with the Levi-Civita/Weitzenböck connection.

The spin connection serves to retain the invariance of the TG field equations under local Lorentz transformations \cite{Golovnev:2017dox}. Hence, for any particular tetrad, the spin connection balances this freedom with different inertial contributions, and can vanish for particular frame choices. Henceforth we consider the Weitzenböck gauge $\omega_{\enspace b\mu}^{a} = 0$. One must ensure to choose a tetrad which obeys this gauge, otherwise one could restrict the theory. This problem is also known as a choice of \textit{good} and \textit{bad} tetrads \cite{Hohmann:2019nat,krvsvsak2019teleparallel}.

The action in TG theories has several cases, the first and easiest is called Teleparallel Equivalent of General Relativity (TEGR). This is very important because the variation of such action with respect to the tetrad field results in completely equivalent to GR dynamical equations \cite{Bahamonde:2019jkf}. The Lagrangian density in this case takes the form $-T+B$. Raising the TEGR action to its $f(T,B)$ gravity extension results in
\begin{equation}\label{fTBaction}
\mathcal{S} = \int d^4x
\bigg[\dfrac{1}{16\pi}f(T,B) + \mathcal{L}_{\rm m}\bigg]e,
\end{equation}
where $e$ is the determinant of the tetrad field $e = \det (e^{A}_{\enskip \mu}) = \sqrt{-g}$. $T$ is the torsion scalar and $B$ is the boundary term denoted, respectively, by
\begin{equation}
T := S_{A}^{\enspace \mu \nu}T_{\enspace \mu \nu}^{A}, \quad B = \frac{2}{e}\partial_{\mu}\Big(e\tensor{T}{^\nu_\nu^\mu}\Big),
\end{equation}
where we use the superpotential and the torsion tensor defined as
\begin{equation}\label{superpotential}
S_{A}^{\enspace \mu \nu}:= \Gamma_{\enspace \mu \nu}^{\sigma} - \mathring{\Gamma}_{\enspace \mu \nu}^{\sigma}-E_{A}^{\enspace \nu}T_{\quad \alpha}^{\alpha \mu}+E_{A}^{\enspace \mu}T_{\quad \alpha}^{\alpha \nu}.
\end{equation}
\begin{equation}\label{torsiontensor}
T_{\enspace \mu\nu}^{\sigma}:=  2\Gamma_{\enspace [\nu \mu]}^{\sigma}.
\end{equation}
$\mathcal{L}_{\rm m}$ represents the Lagrangian for matter and $f=f(T,B)$ is an arbitrary function of $T$ and $B$.

This generalization of TG provides a richer class of models and is quite powerful as it allows to compare TG with $f(\mathring{R})$ models. $f(T,B)$ gravity has also been well-studied \cite{sebas2, capo2, andro, gabriel2, sebas4, matthew, Bahamonde:2016grb, Bahamonde:2020lsm} as a possible extension to TEGR.
Taking a variation of Eq.~(\ref{fTBaction}) with respect to the tetrad field leads to
\begin{widetext}
\begin{eqnarray}\label{fieldFTB}
2\delta_{\nu}^{\lambda}\mathring{\Box}f_{B}-2\mathring{\nabla}^{\lambda}\mathring{\nabla}_{\nu}f_{B}+Bf_{B}\delta_{\nu}^{\lambda}+4\partial_{\mu}(f_{B}+f_{T})S_{\nu}{}^{\mu\lambda}
+4e^{-1}e^{A}{}_{\nu}\partial_{\mu}(eS_{A}{}^{\mu\lambda})f_{T}-4f_{T}T^{\sigma}{}_{\mu\nu}S_{\sigma}{}^{\lambda\mu}-f\delta_{\nu}^{\lambda} = 16\pi\Theta_{\nu}^{\lambda},
\end{eqnarray}
\end{widetext}
where $\mathring{\Box} = \mathring{\nabla}^{\delta}\mathring{\nabla}_{\delta}$, $\Theta^{\alpha \mu}$ is the energy-momentum tensor, and $f_{T}$ and $f_{B}$ refer to the derivative of $f$ with respect to the torsion scalar and boundary term, respectively.

%%%%%%%%%%%%%%%%%%%%%%%%%%%%%%%%
%%%%%%%%%%%%%%%%%%%%%%%%%%%%%%%%

\bigskip

\textit{From Schwarzschild to LTB: The Teleparallel approach.-}
From Eq.~(\ref{ScwLTB}) we use a tetrad which is compatible with the Weitzenböck gauge \cite{Hohmann:2019nat,bahamonde2020general}
\begin{equation}
e^a{}_{\mu}=\left(
\begin{array}{cccc}
1 & 0 & 0 & 0 \\
0 & \frac{Y'\cos\phi \sin\theta}{\sqrt{1-k}}  & Y \cos\phi \cos\theta  & -Y \sin\phi \sin\theta  \\
0 & \frac{Y'\sin\phi \sin\theta}{\sqrt{1-k}}   & Y \sin\phi \cos\theta  & Y \cos\phi \sin\theta \\
0 & \frac{Y'\cos\theta}{\sqrt{1-k}}  & -Y \sin\theta & 0 \\
\end{array}
\right)\label{tetrad}\,.
\end{equation}
With this choice of tetrad and considering a dust source, $\Theta_{ab}=\rho u_au_b$ with $\rho$ the matter--energy density, the non-diagonal components of Eq.~(\ref{fieldFTB}) read
\begin{eqnarray}
\dot{f}'_{B} = \frac{\dot{Y}'f'_{B}}{Y'}+\frac{2\dot{Y}(f'_{B}+f'_{T})}{Y},\label{non-diag1}\\
(-1+\sqrt{1-k})(\dot{f}_{B}+\dot{f}_{T})Y'=\sqrt{1-k}(f'_{B}+f'_{T})\dot{Y},\label{non-diag2}
\end{eqnarray}
while the independent diagonal components are 
\begin{widetext}
\begin{eqnarray}
-\frac{\bigg(\Big[]4\sqrt{1-k}-4\Big]Y'-2k'Y+8M'\bigg)f_{T}}{Y'Y^{2}}-\frac{2[1-k]f''_{B}}{Y'^2}+\frac{\Big(2(1-k)YY''+k'YY'-4\sqrt{1-k}Y'^{2}\Big)f'_{B}}{YY'^{3}}\nonumber\\
-\frac{\Big(-4\dot{Y}YY'-2Y^{2}\dot{Y}'\Big)\dot{f}_{B}}{Y'Y^{2}}+\frac{4\sqrt{1-k}\Big(\sqrt{1-k}-1\Big)f'_{T}}{Y'Y}-Bf_{B}+f=16\pi \rho,\label{eq:eqtt}\\
-\frac{\bigg(\Big(4\sqrt{1-k}-4\Big)Y'-2k'Y+8M'\bigg)f_{T}}{Y'Y^{2}}-\frac{2(1-k)f''_{B}}{Y'^2}+\frac{\Big(2(1-k)YY''+k'YY'-4\sqrt{1-k}Y'^{2}\Big)f'_{B}}{YY'^{3}}\nonumber\\
Bf_{B}+\frac{\bigg(\Big(4\sqrt{1-k}-4\Big)Y'-2k'Y+4M'\bigg)f_{T}}{Y^{2}Y'}-2\ddot{f}_{B}+\frac{4(1-k)f'_{B}}{YY'}-f+\frac{4\dot{Y}\dot{f}_{T}}{Y}=0,\label{eq:eqrr}\\
\frac{2(1-k)f''_{B}}{Y'^2}+Bf_{B}+\Bigg(\frac{4 \sqrt{1-k}-4}{Y^2}+\frac{-2k'Y+4M'}{Y^{2}Y'}\Bigg)f_{T}-2\ddot{f}_{B}-\frac{\Big(2(1-k)YY''+k'YY'-2\sqrt{1-k}Y'^{2}\Big)f'_{B}}{YY'^{3}}\nonumber\\
-f-\frac{2\sqrt{1-k}\Big(\sqrt{1-k}-1\Big)f'_{T}}{Y'Y}+\frac{2(\dot{Y}Y'+Y\dot{Y}')\dot{f}_{T}}{Y'Y}=0.\label{eq:eqthth}
\end{eqnarray}
\end{widetext}

\noindent To find an exact solution to Eq.~(\ref{fieldFTB}) we consider the following ansatz  
\begin{equation}\label{ansatz1}
\dot{f}_{T} = -\dot{f}_{B}, \quad f'_{T} = -f'_{B},
\end{equation}
in Eqs.~(\ref{non-diag1})-(\ref{eq:eqthth}). Eq.~(\ref{non-diag1}) integrates to  
\begin{equation}\label{eqfBr}
f'_{B} = AY'.
\end{equation}
We consider now the spatially flat subcase $k = 0$ of Eq.~(\ref{ScwLTB}) ($k$ is proportional to the 3-dimensional Ricci scalar of constant $\tau$ slices). For this case Eq.~(\ref{rdot}) has a closed solution 
\begin{equation}\label{ansatzonY}
Y = \left(\frac{9}{2}M(t-t_{b}(r))^2\right)^{1/3},
\end{equation}
which reduces the complexity of Eqs.~(\ref{eq:eqtt})-(\ref{eq:eqthth}), through a functional form for $f$ is still needed to obtain an exact TG solution. In particular, the Mixed Power Law Model (MPLM) 
\begin{equation}\label{MPLM}
f(T,B) = \mathcal{F}B^{n}T^{m},
\end{equation}
where $\mathcal{F}$, $n$ and $m$ are free parameters, applied to FLRW metric seems to be a convenient choice complying with viable cosmological scenarios under TG, leading to a power law solution for the scale factor \cite{Bahamonde:2016grb} thus allowing for an evolution that either mimics a phantom energy, or preserves quintessence behavior at high redshifts \cite{Escamilla-Rivera:2019ulu}, admitting as well a continuous transition between cosmological epochs. 

\noindent Substituting Eq.~(\ref{MPLM}) and Eq.~(\ref{ansatzonY}) into Eq.~(\ref{eqfBr}), we obtain
\begin{widetext}
\begin{eqnarray}
AY^{3/2}M'^{3}+C_{nm}\left(\frac{MM'}{-M\sqrt{2}Y^{\frac{3}{2}}\sqrt{M}t'_{B}+\frac{1}{3}M'Y^{3}}\right)^{m+n}(m+n-1)n\mathcal{F}\Xi=3AM^{3/2}t'_{B}\sqrt{2}M'^{2}, \label{resk0mn}
\end{eqnarray}
\end{widetext}
where $C_{nm}=2^{2m+n-1}3^{n+1}$, and
\begin{widetext}
\begin{equation}
\Xi = t'_{B}M^{2}\sqrt{Y}M't'_{B}+\frac{1}{3}Y^{2}\sqrt{2}\Big(t'_{B}M^{\frac{3}{2}}M''-t''_{B}M^{\frac{3}{2}}M'-2M^{\frac{1}{2}}M'^{2}t'_{B}\Big).
\end{equation}
\end{widetext}
Since the RHS of Eq.~(\ref{resk0mn}) depends only on $r$, while the LHS is a function of $t,r$, the following solutions arise:
\begin{enumerate}[label=(\alph*)]
\item $n=0$ and $A=0$,
\item $m+n = 1$ and $A = 0$,
\end{enumerate}
with case 1 being the unique TEGR solution, leaving case 2 as a non-trivial and non-static exact inhomogeneous TG solution. This is an exact solution since  Eqs.~(\ref{eq:eqrr})-(\ref{ansatz1}) are immediately satisfied for $\dot Y\ne 0$ and with no additional restrictions on the constants $m$ and $n$. 

%%%%%%%%%%%%%%%%%%%%%%%%%%%%%%%%%%%%%%%%%%%%%%%
%%%%%%%%%%%%%%%%%%%%%%%%%%%%%%%%%%%%%%%%%%%%%%%

%%%%%%%%%%%%%%%%%%%%%%%%%%%%%%%%%%%%%%%%%%%%
%%%%%%%%%%%%%%%%%%%%%%%%%%%%%%%%%%%%%%%%%%%%
\rule{85mm}{0.7mm}
%%%%%%%%%%%%%%%%%%%%%%%%%%%%%%%%%%%%%%%%%%%%
%%%%%%%%%%%%%%%%%%%%%%%%%%%%%%%%%%%%%%%%%%%%
\textit{Conclusions}.-
We have derived and presented in this paper the first inhomogeneous non-static exact solution in $f(T,B)$ gravity for the LTB metric. We remark that \textit{(i)} this is an exact solution for a form of $f(T,B)$ that is compatible with a viable cosmological scenario and \textit{(ii)} the solution corresponds to a spherically symmetric dust source that satisfies the same evolution equation for the area radius Eq.~(\ref{rdot}) as its equivalent LTB models in GR. 

While we are now presenting a straightforward  solution, we believe that it is the first step that should serve to motivate the search for inhomogeneous models in extended gravity theories that should be potentially useful for cosmological and astrophysical applications. Given the utility of LTB models in GR, the next step is to consider the general class of models (with $k\neq 0$, $\Lambda >0$) under a TG approach. An attempt to consider avoidance of specific ansatzes or even considering a modified LTB metric should also be done. This task is being currently undertaken and its results will be reported elsewhere.\\   

%%%%%%%%%%%%%%%%%%%%%%%%%%%%%%%%%%%%%%%%%%%%
%%%%%%%%%%%%%%%%%%%%%%%%%%%%%%%%%%%%%%%%%%%%

\begin{acknowledgments}
CE-R acknowledges the Royal Astronomical Society as FRAS 10147. CE-R and A.A are supported by PAPIIT-UNAM Project IA100220 and would like to acknowledge networking support by the COST Action CA18108.
SN acknowledges financial support from SEP–CONACYT postgraduate grants program and RAS acknowledges support from PAPIIT-DGAPA RR107015. We thank Jackson Levi Said for useful and enlightening discussions.

\end{acknowledgments}

%%%%%%%%%%%%%%%%%%%%%%%%%%%%%%%%%%%%%%%%%%%%
%%%%%%%%%%%%%%%%%%%%%%%%%%%%%%%%%%%%%%%%%%%%

\bibliographystyle{unsrt}
\bibliography{references}
\end{document}